\documentclass[aps,floatfix,twocolumn]{revtex4}
\usepackage[pdftex]{graphicx}
\usepackage{amsmath}
\usepackage{hyperref}
\usepackage{latexsym}
\usepackage{color}
\usepackage{amssymb}
\usepackage{amsmath}

\usepackage [latin1]{inputenc}

\usepackage{bm}
\usepackage{graphicx}
\usepackage{siunitx}
\usepackage{amsmath,amssymb}
\usepackage{braket}
\usepackage[final]{microtype}

\begin{document}

\title{Electro-association of ultracold dipolar molecules \\
 into tetramer field-linked states}

\author{Goulven Qu{\'e}m{\'e}ner}
\affiliation{Universit\'{e} Paris-Saclay, CNRS, Laboratoire Aim\'{e} Cotton, 91405 Orsay, France}
\author{John L. Bohn}
\affiliation{JILA, NIST, and Department of Physics, University of Colorado, Boulder, Colorado 80309-0440, USA}
\author{James F. E. Croft}
\affiliation{The Dodd-Walls Centre for Photonic and Quantum Technologies, Dunedin 9054, New Zealand}
\affiliation{Department of Physics, University of Otago, Dunedin 9054, New Zealand}

\begin{abstract}
The presence of electric or microwave fields can modify the long-range forces between ultracold dipolar molecules in such a way as to engineer weakly-bound states of molecule pairs.  These so-called field-linked states [Avdeenkov et al., Phys. Rev. Lett. 90, 043006 (2003), Lassabli{\`e}re et al., Phys. Rev. Lett. 121, 163402 (2018)], in which the separation between the two bound molecules can be orders of magnitude larger than the molecules themselves, have been observed as resonances in scattering experiments [Chen et al., Nature 614, 59 (2023)].  Here, we propose to use them as tools for the assembly of weakly-bound tetramer molecules, by means of ramping an electric field, the electric-field analog of magneto-association in atoms. This ability would present new possibilities for constructing ultracold polyatomic molecules.
\end{abstract}

\maketitle

Field-linked states (FLS) appear in carefully engineered long-range wells,
of two colliding ultracold molecules.
The wells are created by applying an external field
when the molecules are in a specific initial state.
They were first predicted in 2003~\cite{Avdeenkov_PRL_90_043006_2003,Avdeenkov_PRA_66_052718_2002,Avdeenkov_PRA_69_012710_2004,Avdeenkov_PRA_71_022706_2005}
using a static electric field.
In 2018~\cite{Lassabliere_PRL_121_163402_2018}, a second type of FLS
were predicted using a blue-detuned microwave field
on ground-state ultracold dipolar molecules.
In 2023, resonances due
to this second type of ``microwave'' FLS were experimentally observed~\cite{Chen_N_614_59_2023}.
This discovery opens a wide range of possibilities that have so far been lacking for ultracold molecules,
the most important being the ability to control the molecule-molecule
scattering length at will~\cite{Lassabliere_PRL_121_163402_2018},
from positive to negative, from a small to a large value.
This control is key in order to reach the regime needed to explore strongly correlated states of matter
involving molecular Bose-Einstein Condenstates (BEC) and degenerate Fermi gases (DFG)
\cite{DeMarco_S_363_853_2019,Valtolina_N_588_239_2020,Schindewolf_N_607_677_2022},
made possible by the electric dipole moment of the molecules.
For example, studying the BEC-BCS cross-over
\cite{Regal_PRL_92_040403_2004,Bartenstein_PRL_92_120401_2004,Zwierlein_PRL_92_120403_2004,
Bourdel_PRL_93_050401_2004} could now be within reach for fermionic dipolar molecules,
in a regime where few-body dipolar physics could come into play~\cite{Wang_PRL_107_233201_2011}.
Similarly, studies of Efimov physics \cite{Efimov_PLB_33_563_1970,Kraemer_N_440_315_2006,Braaten_PRep_428_259_2006}
with bosonic dipolar molecules~\cite{Wang_PRL_106_233201_2011}
could also be made possible.
Finally, one could tune both the scattering length and the dipolar length in an electric dipolar molecular BEC
to explore density profiles with supersolid properties~\cite{Schmidt_PRR_4_013235_2022},
complementing those already observed for magnetic dipolar atomic BEC's~\cite{Tanzi_PRL_122_130405_2019,Chomaz_PRX_9_021012_2019,Bottcher_PRX_9_011051_2019}.

The ability to use a magnetic field to control the scattering length via Fano--Feshbach resonances,
as is routine for atoms~\cite{Kohler_RMP_78_1311_2006,Chin_RMP_82_1225_2010},
seems to rarely occur for bi-alkali molecules.
Except for the lightest bi-alkali molecule of LiNa in the triplet electronic state
\cite{Park_N_614_54_2023}, it has proven somewhat difficult to find well-isolated resonances
in heavy molecule-molecule systems in their ground electronic singlet states,
due to their high density of states
(see for example \cite{Croft_PRA_102_033306_2020, Croft_PRA_107_023304_2023,Bause_JPCA_127_729_2023}).
The existence of FLS could circumvent this difficulty and present an alternative way
to both shield the molecules against losses and tune the scattering length.

In addition, we note the power of magnetic field ramps 
to produce bound atomic dimers across Fano--Feshbach resonances
\cite{Kohler_RMP_78_1311_2006,Chin_RMP_82_1225_2010}.
Here we explore the feasibility of associating two ultracold dipolar molecules
into a long-range, field-linked tetramer state, using an electric field.  
Such a technique promises to be a powerful starting point 
for the controlled assembly of polyatomic molecules, starting from diatomic ingredients.
By ramping the electric field of a microwave,
we show that electro-association (EA) of two colliding dipolar molecules can be achieved using experimentally realistic ramps and traps.
We discuss the conditions under which EA is applicable, 
particularly 
in terms of the initial population distribution 
of the molecular gas 
and the bosonic/fermionic character of the molecules.

\begin{figure}[t]
\begin{center}
\includegraphics*[width=8.5cm,trim=0.5cm 0cm 0cm 0cm]{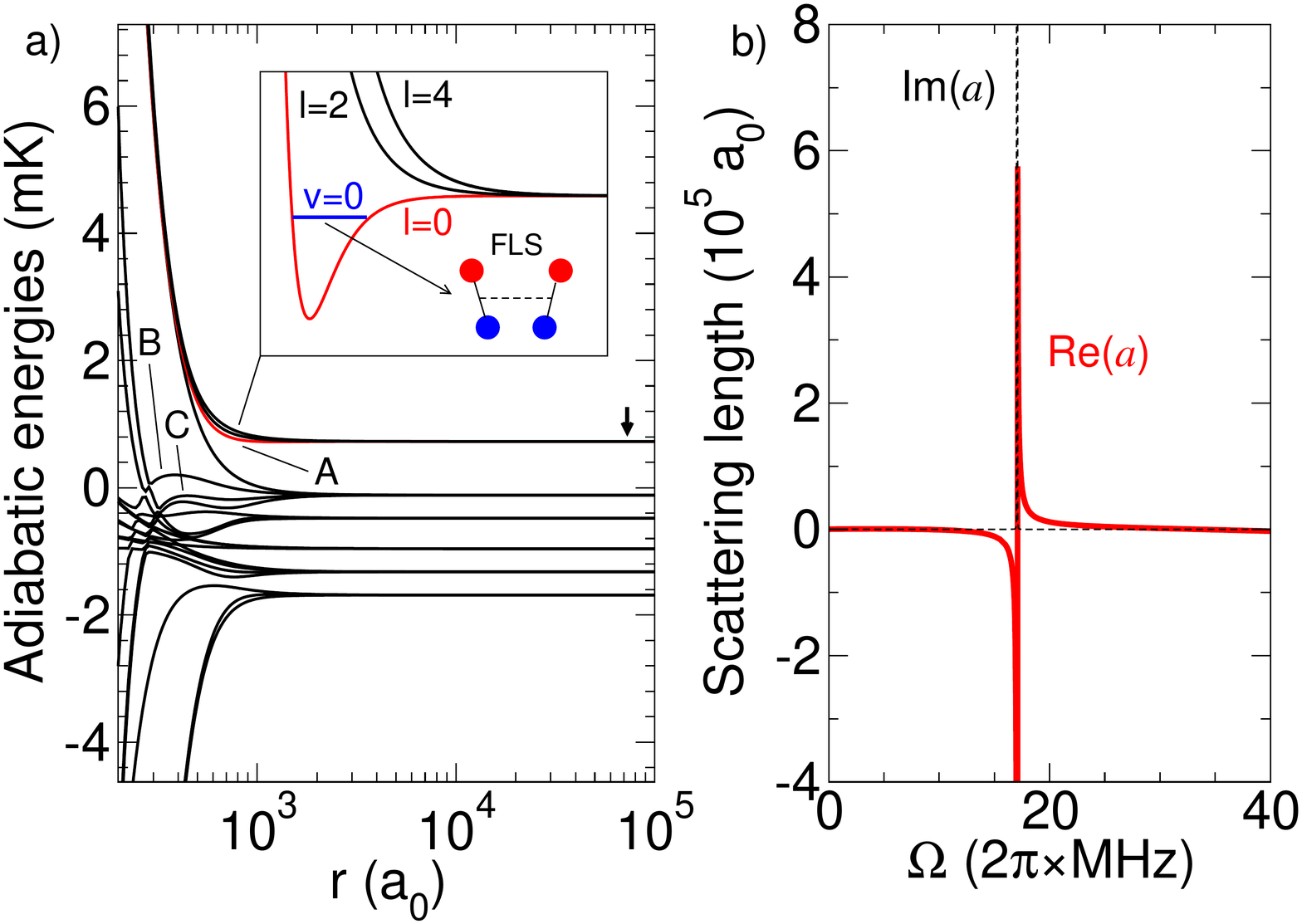} 
\caption{ 
a) Potentials of two colliding molecules as a function of the inter-molecular separation $r$, for $\Delta = 10$~MHz and $\Omega = 2\pi \times 40$~ MHz. The initial colliding state is indicated by an arrow. The inset shows a zoom of the long-range well in red that can carry a field-linked state (FLS). The letters A, B, C label respectively the entrance channel and two scattering channels into which the FLS can decay.
b) Scattering length as a function of $\Omega$ for $\Delta=10$~MHz. 
}
\label{FIG-SPAG-SCTL}
\end{center}
\end{figure}

We first consider an ultracold gas of dipolar bosonic bi-alkali
molecules in their ground electronic, vibrational and rotational state,
taking $^{23}$Na$^{87}$Rb as a representative example \cite{Guo_PRL_116_205303_2016,Christakis_N_614_64_2023}. 
Many other molecules have been created under similar conditions
\cite{Ni_S_322_231_2008,Aikawa_PRL_105_203001_2010,Takekoshi_PRL_113_205301_2014,
Molony_PRL_113_255301_2014,Park_PRL_114_205302_2015,Rvachov_PRL_119_143001_2017,
Liu_S_360_900_2018,
Anderegg_S_365_1156_2019,Voges_PRL_125_083401_2020,Bause_PRR_3_033013_2021,Stevenson_arXiv_2206_00652_2022,
Ruttley_arXiv_2302_07296_2023},
and can also be considered without loss of generality.
When a circularly polarized microwave field is applied on molecules in their ground rotational state $j=0$, 
and if the microwave is slightly blue-detuned with respect to the first excited state $j=1$,
it was proposed \cite{Lassabliere_PRL_121_163402_2018,Karman_PRL_121_163401_2018}
and observed \cite{Anderegg_S_373_779_2021,Schindewolf_N_607_677_2022,
Bigagli_arXiv_2303_16845_2023,Lin_arXiv_2304_08312_2023}
that a shielding against collisional losses takes place at appropriate values of
the Rabi frequency $\Omega = d E / \hbar$,
enabling long-lived molecules. 
$\Omega$ is proportional to the electric field $E$ of the microwave and the
electric dipole moment $d$ of the molecule.

Potentials that lead to shielding of two colliding NaRb molecules
are shown in Fig.~\ref{FIG-SPAG-SCTL}-a where they
are plotted as a function of the inter-molecular separation $r$,
for a detuning $\Delta = 10$~MHz and $\Omega = 2\pi \times 40$~MHz.
The initial colliding state (entrance channel) is indicated by a black arrow, corresponding to two molecules
in $j=0$. In this situation, the potentials of the entrance channel become repulsive at $r \approx 10^3 a_0$ when dressed by the microwave. 
This prevents the molecules from getting too close to each other and being lost due to any short-range loss mechanism.
The red curve corresponds to the entrance channel for the $l=0$ partial wave
and it is shown in the inset of the figure 
(we emphasize that the radial coordinate axis in Fig.~\ref{FIG-SPAG-SCTL} is on a logarithmic scale highlighting the extremely long range nature of these states).
The two other black curves of the inset correspond to an entrance channel with $l=2$ and $l=4$.
The values of $l$ are even as we are considering indistinguishable bosonic molecules.
The $l=0$ curve is attractive as expected for an $s$-wave
and this creates a long-range well that can hold a FLS~\cite{Avdeenkov_PRL_90_043006_2003,Avdeenkov_PRA_66_052718_2002,Avdeenkov_PRA_69_012710_2004,Avdeenkov_PRA_71_022706_2005},
represented in blue and characterized by a quantum number $v=0$.
The $l=2$ and $l=4$ curves are repulsive due to the presence of the associated
centrifugal barriers and do not form wells.
As such in the following we will solely focus on the $s$-wave curve, in which the FLS exists.
Presumably, similar conclusions on the control of the scattering length
and the existence of FLS also apply to optical shielding~\cite{Xie_PRL_125_153202_2020,Karam_arXiv_2211_08950_2022,Napolitano_PRA_55_1191_1997},
providing an alternative route for experimental investigations.

The presence of a FLS enables the tunability of the scattering length, as can be seen
in Fig.~\ref{FIG-SPAG-SCTL}-b.
Generally, the scattering length $a$ is a complex quantity \cite{Balakrishnan_CPL_280_5_1997,Hutson_NJP_9_152_2007}
expressed as
$a = \text{Re}(a) - i \text{Im}(a)$ with $\text{Im}(a) \ge 0$ .
When $\Omega$ is varied from 0 to $2\pi \times$ 40~MHz,
$\text{Re}(a)$ diverges and changes sign at $\Omega \simeq 2\pi \times 17$~MHz,
from a big negative value to a big positive value.
This is related to the appearance at threshold of a new bound state in the long-range well,
and is the origin of the FLS seen at $2\pi \times$ 40~MHz in Fig.~\ref{FIG-SPAG-SCTL}-a.
Such a variation of the scattering length is reminiscent of magneto-association (MA),
where now the electric field (via $\Omega$) plays the role of the magnetic field.
As such one can use a similar technique to associate two molecules,
hence the name ``electro-association''.

\begin{figure}[t]
\begin{center}
\includegraphics*[width=8.5cm,trim=0.5cm 0cm 0cm 0cm]{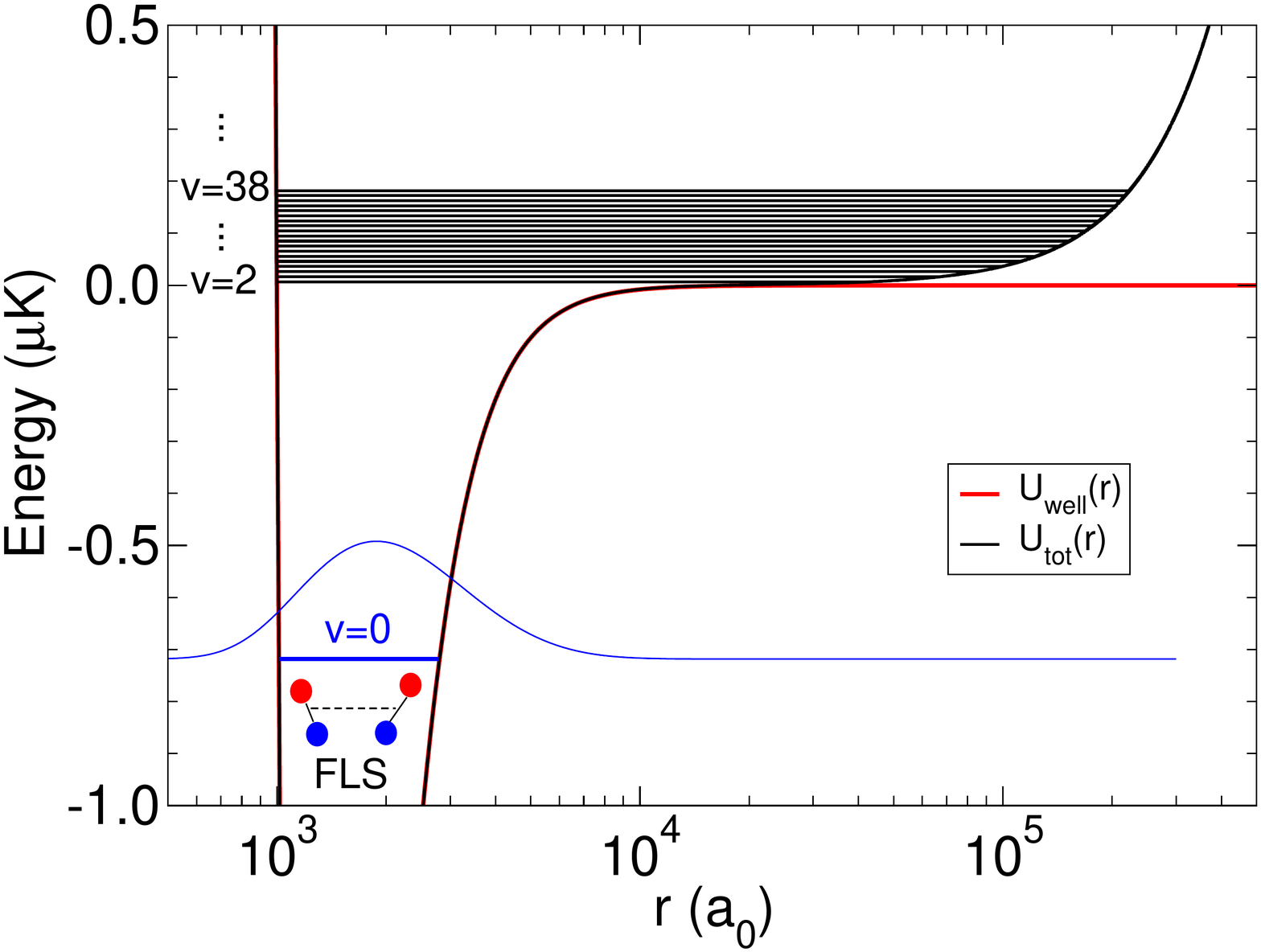} 
\caption{
Energies $E_v$ of the eigenstates (blue and black horizontal lines) 
in the potential energy $U_\text{tot}$ (black curve) 
comprised of 
the long-range well $U_\text{well}$
of Fig.~\ref{FIG-SPAG-SCTL}-a (red curve) 
and the trapping potential of the molecules $U_\text{trap}$, for $\nu = 100$~Hz and $l=0$. 
The FLS is plotted in blue with $v=0$ and negative energy, 
the HOS are plotted in black with  
$v=2, \ldots, 38$ and positive energies.
This is an example for $\Delta = 10$~MHz and $\Omega = 2\pi \times 40$~ MHz.
}
\label{FIG-POT-NRG}
\end{center}
\end{figure}

We now consider the effect of a trapping potential on the molecular collisions,
which causes the relative motion between molecules to be quantized~\cite{Mies_PRA_61_022721_2000,Tiesinga_PRA_61_063416_2000}.
In the experiments cited above,
the molecules are generally trapped in an Optical Dipole Trap (ODT), that can be well represented by a harmonic potential.
To simplify the study, we will consider a spherical ODT with a unique frequency $\nu = 100$~Hz.
The trapping potential is then given by $U_\text{trap}(r_i) = m \, \omega^2 r_i^2/2$, with $\omega = 2 \pi \, \nu$, $m$ the mass of a molecule and $r_i$ the position of molecule $i$, with $i=1,2$.
To treat the collision between two molecules we use coordinates for the relative motion in the center-of-mass frame~\cite{Quemener_PRA_83_012705_2011,Mies_PRA_61_022721_2000,Tiesinga_PRA_61_063416_2000},
as the center-of mass is not involved in the process.
One can then show that, for identical molecules seeing the same frequency $\nu$ of the trap,
the problem can be recast using a trapping potential for the relative motion given by $U_\text{trap}(r) = \mu \, \omega^2 r^2/2$, where $\mu = m/2$ is the reduced mass of the 
two molecules and $r$ the inter-molecular separation.
As the center-of-mass motion is independent and uncoupled from the relative motion, it is not taken into account in the following.

\begin{figure}[t]
\begin{center}
\includegraphics*[width=8.5cm,trim=0cm 0cm 0cm 0cm]{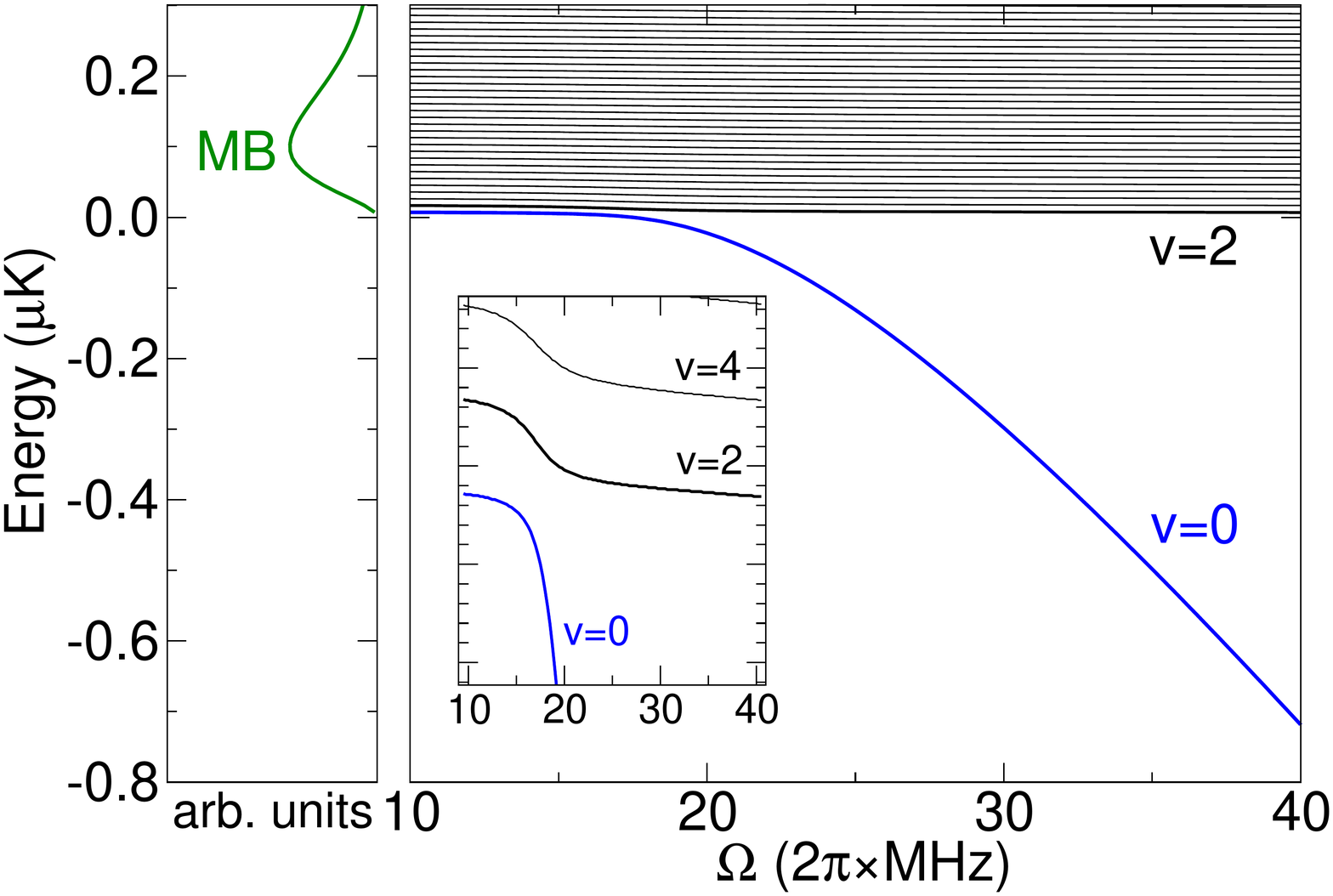} 
\caption{
Energies $E_v$ of many eigenstates shown in Fig.~\ref{FIG-POT-NRG} for $l=0$, 
now as a function of $\Omega$. 
The inset shows a blow-up of the avoided 
crossing between the state $v=0$ and $v=2$. 
The left panel shows a typical initial Maxwell-Boltzmann 
distribution of an individual molecule 
for $T=50$~nK. 
}
\label{FIG-NRG-DIST}
\end{center}
\end{figure}

The overall potential $U_\text{tot}(r)$ seen by two ultracold molecules in $j=0$ is then composed of the long-range well
$U_\text{well}(r)$, shown in red in Fig.~\ref{FIG-SPAG-SCTL}-a and the trapping potential $U_\text{trap}(r)$.
This potential is plotted in Fig.~\ref{FIG-POT-NRG} and shows its spatial extent.
The discrete energies of the molecules are shown in blue for the lowest state along with the corresponding wavefunction, and in black for the higher states.
The energies were computed numerically by solving the radial Schr{\"o}dinger equation for a given value of $\Omega$,
for the Hamiltonian $H(r) = -(\hbar^2/2 \mu) d^2/dr^2 + U_\text{tot}(r)$ for $l=0$.
Each energy and wavefunction is characterized by a quantum number $k=0,1,2, \ldots$ 
that counts the number of the state.
The energies $E_k$ and wavefunctions $f_k(r)$ are found when the outward and inward log-derivative of the wavefunctions match
\cite{Johnson_JCP_67_4086_1977,Hutson_CPC_84_1_1994,Thornley_JCP_101_5578_1994}. 
For low values of $\Omega$ when the long-range well is not too deep, the energies tend to those of the spherical harmonic oscillator given by $E_v^{\text{ho}} = \hbar \omega (v + 3/2)$, with 
$v=2k+l$ \cite{Mies_PRA_61_022721_2000,Tiesinga_PRA_61_063416_2000}. 
A state $v$ has a degeneracy $g_v = (v+1)(v+2)/2$.
$l$ is the quantum number associated with the orbital angular momentum of the relative motion in the spherical harmonic potential and is the same value 
as the partial wave introduced above when two particles collide. 
For indistinguishable bosonic molecules, $v$ takes even values as $l$ are even.
If we would have considered indistinguishable fermionic molecules, $v$ would have taken odd values as $l$ 
would have been odd.
In our study, we consider just $l=0$ as higher values $l=2,4,\ldots$ are not relevant as they do not provide conditions
for the existence of FLS.
Then, $v \equiv 2k$ and $v=0,2,\ldots$. For fermions, we would have had $v=1,3,\ldots$.
Typically at $\nu = 100$~Hz, $E_0^{\text{ho}} \simeq 7.2$~nK, $E_2^{\text{ho}} \simeq 16.8$~nK,  etc \ldots.
We will refer to an eigenstate of positive energy as an harmonic oscillator state (HOS)
and to an eigenstate of negative energy as a FLS. 
For each value of $\Omega$ at $\Delta = 10$~MHz, we compute the energy $E_v$ of the eigenstates. This is shown in Fig.~\ref{FIG-NRG-DIST}. 
The lowest state $v=0$ is plotted in blue. At low values of $\Omega$, 
$E_0 \simeq E^\text{ho}_0$, no FLS exists, 
the energy being simply the one of the lowest HOS. 
When $\Omega$ increases, the energy $E_0$ decreases and become negative
around the value $\Omega = 2\pi \times 17$~MHz. Then the lowest HOS transfers smoothly to the FLS. Not surprisingly, this occurs at the same value of $\Omega$ where the scattering length diverges as seen in Fig.~\ref{FIG-SPAG-SCTL}.

These results are very reminiscent of trapping potentials for 
MA~\cite{Mies_PRA_61_022721_2000,Tiesinga_PRA_61_063416_2000},
even though the physical principle behind the curves is very different.
Most notably only one potential energy curve is sufficient for EA while two are needed for MA.
The idea behind electro-association is to pass smoothly from the lowest HOS,
initially populated by the free well separated molecules,
to the bound tetramer FLS, by ramping the value of $\Omega$ as a function of time via the electric field of the microwave.
Solving the time-dependent Schr{\"o}dinger equation including the contribution of the 
decay lifetimes of the diatomic molecules or the tetramers  yields 
the time evolution of the probability $P_{v v'}$ \cite{SM}
to start in a state $v$ and end up in a state $v'$ 
for the $l=0$ curve,
for a given ramp noted $R = d\Omega/dt$ with $R$ expressed here 
in $ 2\pi \times$MHz/ms.

\begin{figure}[h]
\begin{center}
\includegraphics*[width=8.5cm,trim=0cm 0cm 0cm 0cm]{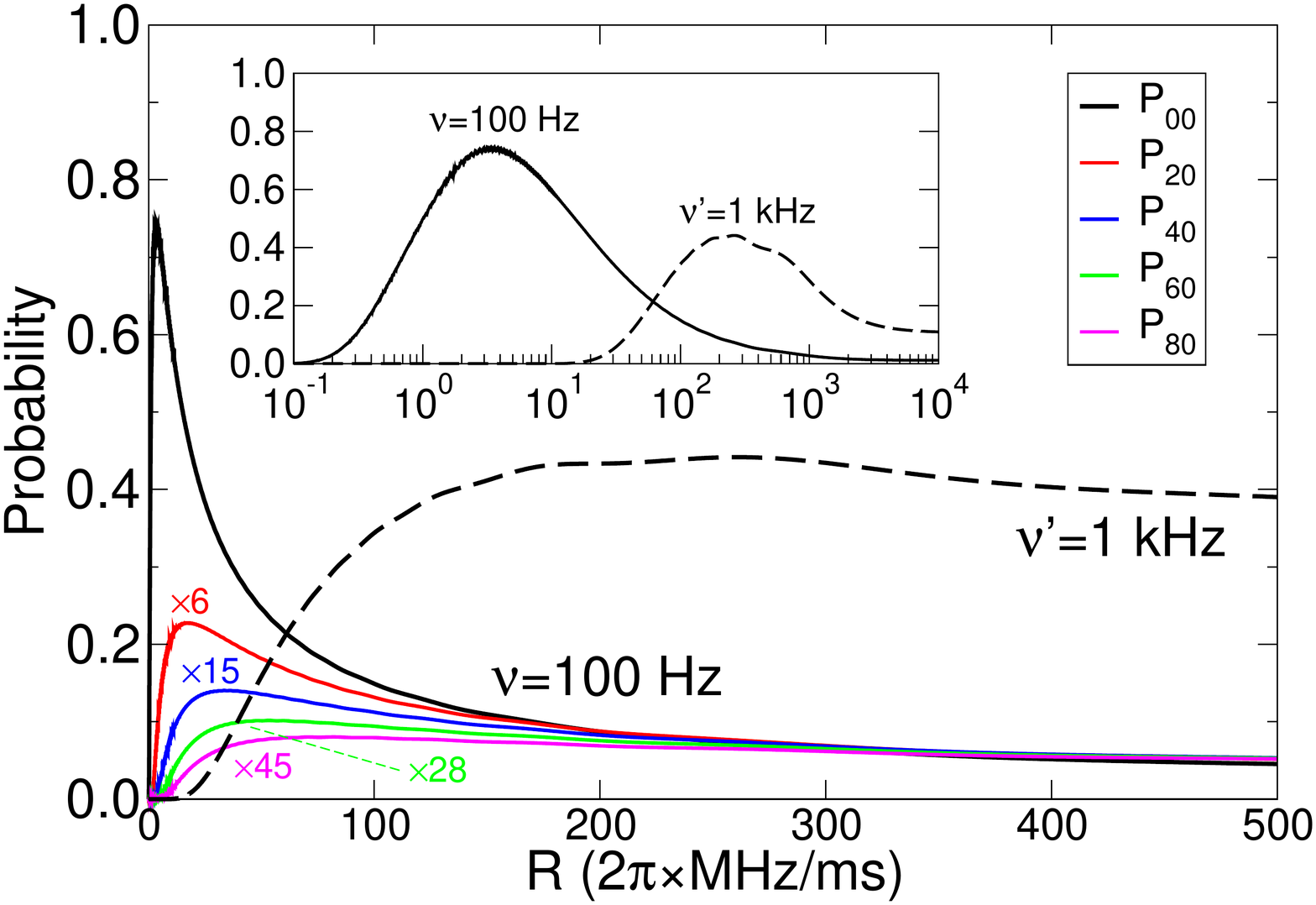} 
\caption{
Probability $P_{0 0}$ to start in the state $v=0$ (representing two trapped molecules) at $\Omega= 2\pi \times 10$~MHz and finish
in the state $v'=0$ (representing a FLS) at $\Omega= 2\pi \times 40$~MHz, for trap frequencies of 
$\nu = 100$~Hz (black solid line)
and $\nu' = 1$~kHz (black dashed line). 
This is plotted as a function of the applied ramp $R$. 
The inset shows the same figure using a logarithmic scale.
The lower curves show the probabilities $P_{v0}$ at $\Omega= 2\pi \times 40$~MHz for $\nu = 100$~Hz,
with $v=2,4,6,8$ shown as red, blue, green and pink lines respectively. 
For visibility in the figure, 
we have multiplied them by $6, 15, 28, 45$ respectively.
}
\label{FIG-RAMP}
\end{center}
\end{figure}

For illustrative purposes we consider a ramp from $\Omega= 2\pi \times 10$~MHz to
$\Omega= 2\pi \times 40$~MHz.
The probability $P_{0 0}$ to start in $v=0$ and end up in $v'=0$
is shown as a black solid line
in Fig~\ref{FIG-RAMP},
as a function of the value of the ramp $R$, for $\nu = 100$~Hz.
We can see that the probability reaches a maximum for a ramp of 
 $R \simeq 2\pi \times$ 3.4~MHz/ms, corresponding to a duration of 8.8~ms
for which an association probability of 75~$\%$ can be reached.  
This shows that one can have relatively 
efficient EA of two molecules 
initially in the lowest HOS of the relative motion 
$v=0$ into a long-range tetramer
under realistic experimental conditions. 
The range of application is however limited (see inset). 
One needs to remain in the range
$R = 2\pi \times$ [1--10] MHz/ms, to 
get probabilities above 50~$\%$.
The optimal ramp is a balance between two competing 
mechanisms.
If the ramp is too slow ($R < 2\pi \times$ 1 MHz/ms), 
then collisions lead to molecule loss during the ramp, and the final yield of FLS will be small.  
On the other hand, if the ramp is too fast ($R > 2\pi \times$ 10 MHz/ms), 
the molecules are unable to adiabatically transfer into the ground 
state $v^{\prime}=0$, with the result that higher trap states are populated, rather than the FLS.

In Fig.~\ref{FIG-RAMP},
we also show the probabilities $P_{v0}$ 
to start in the $l=0$ curve of a 
state $v=2,4,6,8$ of the HOS and to end up in $v'=0$, that is the FLS.
We see that $P_{00}$ is clearly always higher than the other probabilities $P_{v0}$ with $v>0$, whatever the choice of ramp, the other probabilities being less than 4~\%.
Besides the criterion of appropriate ramps, this shows
that EA is also most efficient 
when the relative motion of bosonic molecules occupy the lowest HOS $v=0$.
The same conclusion will also hold for fermionic molecules
except that the lowest HOS would be $v=1$.
In the higher states, 
the probabilities become very small and no real chance is given to form a FLS.

As such the initial distribution will strongly affect the efficiency of the
formation of a FLS.
First we take the example of a initial Maxwell-Boltzmann distribution
for indistinguishable bosonic molecules,
for which the population of a state $v_i$ 
for an individual molecule $i=1,2$
is given by
$A \, g_{v_i} \, \exp(- E_{v_i}^\text{ho}/k_BT)$,
where $A$ is the normalization factor and $g_{v_i}$ the degeneracy of state $v_i$.
We show an example distribution in the left panel of Fig.~\ref{FIG-NRG-DIST}
as a function of the discrete energy of an individual trapped molecule
for a typical temperature of $T=50$~nK.
The distribution shows that the molecules populate
dominantly high excited states, 
namely $v_1, v_2 \simeq 20$ at that temperature. 
To express the distribution of two individual molecules $i=1,2$
in the relative and center-of-mass representation
in which the previously derived results hold, we have to use the composition of the harmonic oscillator functions
derived for example in Appendix C of \cite{Quemener_PRA_83_012705_2011}.
It is shown that, for any symmetric constructions
in the individual representation
between the $v_1$ and $v_2$ states,
there will be always a $v=0$ component in the relative one.
For a thermal gas of indistinguishable fermionic molecules 
it is also shown in Appendix C of
\cite{Quemener_PRA_83_012705_2011},
that for any anti-symmetric construction
of $v_1$ and $v_2$ states, there will be always a corresponding $v=1$ component
with the difference with the bosonic case now
that one has to have $v_1 \ne v_2$ 
(by definition of anti-symmetric states). 
Then, any vibrational state populated at a given temperature $T$ 
in Fig.~\ref{FIG-NRG-DIST}
will have at least some projection onto a $v=0$ state for bosons
(or a $v=1$ state for fermions)
and can contribute to EA.
This statement is in agreement with former results obtained 
in the context of MA of ultracold bosonic and fermionic atoms into weakly-bound molecules \cite{Hodby_PRL_94_120402_2005}.

We finally consider starting from a BEC.
In this case, 
all the molecules start in the lowest state $v_1=v_2=0$
which is equivalent to occupying  
the $v=0$ state. 
To illustrate the feasibility of such an example, one has 
to take into account the many-body interactions in a BEC \cite{Mies_PRA_61_022721_2000,Tiesinga_PRA_61_063416_2000,Borca_NJP_5_111_2003,Goral_JPBAMOP_37_3457_2004}.
Here we follow the work of \cite{Borca_NJP_5_111_2003}
where the authors showed that 
one can rescale the trap frequency to a new effective one, $\nu' = (\hbar \, / \, 2m) \, (n/2)^{2/3}$.
For example, a typical BEC density of $n = 1.3 \ 10^{13}$ molecules/cm$^3$ yields 
an effective confinement of $\nu' = 1$~kHz for NaRb.
As shown in Fig.~\ref{FIG-RAMP}, 
the probability $P_{00}$ at such frequency
now reaches a maximum for a ramp of
$R \simeq 2\pi \times$ 260 MHz/ms, corresponding to a duration of 0.11 ms. 
The  optimal association probability is only 44~$\%$ 
which is lower than for $\nu = 100$~Hz. 
This is because in a BEC the density is much higher
and collisions are therefore 
more frequent decreasing the overall efficiency of EA.
As such the optimal ramp has to be faster
compared to the $\nu = 100$~Hz case.
This can be afforded in this case as the rescaled trap frequency $\nu'$ is higher than $\nu$. The avoided crossing seen in Fig.~\ref{FIG-NRG-DIST} will now be more avoided, still allowing 
a high transfer to the FLSs even at such a large ramp.
In the situation of a molecular DFG 
\cite{DeMarco_S_363_853_2019,Valtolina_N_588_239_2020,Schindewolf_N_607_677_2022} 
one could also consider starting with distinguishable fermions as was done in the context of MA,
provided that the microwave shielding and 
the existence of FLS still hold for that case.
If that is the case, $l=0$ and $v=0$ would be allowed,
as well as symmetric constructions of $v_1$ and $v_2$ states, increasing
the number of combinations of $v_1$ and $v_2$ allowed, and the conversion rate for the fermionic case.

In conclusion, we have shown that
electro-association of two ultracold dipolar molecules 
into a tetramer field-linked state is possible
using realistic ramps of the electric field of a microwave,
when the molecules are 
in the ground state of their relative motion.
For typical thermal gases with typical densities
and typical harmonic traps used in experiments, 
any vibrational state of a Maxwell-Boltzmann
distribution can contribute to EA.
An ideal situation occurs for typical BEC conditions
where most of the molecules
are directly in the ground state of their relative motion.
But in both cases, EA efficiency is affected by collisional losses.
To further prevent such collisions, 
one could consider starting with exactly two molecules
in the ground state of a strong confined trap
such as an optical lattice \cite{DeMiranda_NP_7_502_2011,Chotia_PRL_108_080405_2012,Yan_N_501_521_2013,DeMarco_S_363_853_2019,
Christakis_N_614_64_2023}
or optical tweezers \cite{Liu_S_360_900_2018,Anderegg_S_365_1156_2019,Anderegg_S_373_779_2021,
Kaufman_NP_17_1324_2021,Ruttley_arXiv_2302_07296_2023}.
This could be an ideal set up for future prospects of transferring
such long-range bound excited states to ground state tetramers,
as was done in the context of formation of diatomic molecules
using stimulated Raman adiabatic processes \cite{Bergmann_RMP_70_1003_1998,Bergmann_JCP_142_170901_2015,Vitanov_RMP_89_015006_2017}.
This constitutes
an alternative to photo-association proposals \cite{Lepers_PRA_88_032709_2013,Perez-Rios_PRL_115_073201_2015,Gacesa_PRR_3_023163_2021}
and could open the way for a piece-by-piece production type of bigger, ultracold molecules.
\\

J.L.B acknowledges that
this material is based upon work supported by the National Science Foundation under Grant No. PHY 1734006.
J.F.E.C acknowledges support by the Marsden Fund of New Zealand (Contract No. 19-UOO-130) as well as a Dodd-Walls Fellowship.

\section*{Supplemental material}

\subsection{Dynamics of the ramp using a time-dependent Schr{\"o}dinger equation}

We consider the $l=0$ channel containing the FLS
described by the potential $U_\text{tot}(r) = U_\text{well}(r) + U_\text{trap}(r)$.
When a ramp of $\Omega(t)$ is applied, 
the overall potential now depends parametrically on time via $\Omega$,
which we denote $U_\text{tot}^{\Omega}(r)$.
One can obtain the corresponding wavefunctions 
and energies 
obeying 
\begin{eqnarray}
H^{\Omega}(r) f_v^{\Omega}(r)  = E_v^{\Omega} \, f_v^{\Omega}(r)
\end{eqnarray}
with $H^{\Omega}(r) = -(\hbar^2/2 \mu) d^2/dr^2 + U_\text{tot}^{\Omega}(r)$ 
and where $H, f_v, E_v$ also now depend parametrically on time via $\Omega$.
The time-dependent wave function of the two colliding molecules is expanded in this parametric basis 
\begin{eqnarray} \label{TDWF}
\Psi(t) = \sum_{v=0}^{38} c_v(t) \, f_v^{\Omega}(r) \,
\exp(-i E_v^{\Omega} t / \hbar) .
\end{eqnarray}
We consider the first 20 
states plotted in Fig.~\ref{FIG-NRG-DIST} from $v=0$ to $v=38$, which is sufficient to converge the results.
At $t=0$, we consider the molecules starting in $v=0$ at 
$\Omega= 2\pi \times 10$~MHz.
The reason why we choose this value is that for lower value of $\Omega$, the red curve is not yet repulsive enough to protect the molecules from loss, so the ramp has at least 
to start from this value.
The ramping process will end for example at $\Omega= 2\pi \times 40$~MHz.
By solving the time-dependent Schr{\"o}dinger equation
using Eq.~\eqref{TDWF} 
with the initial conditions $c_0(0)=1$ and $c_{v \ne 0}(0)=0$, we can monitor the evolution
of the complex coefficients $c_v(t)$ during the ramp. These coefficients obey the set of coupled differential equations 
\begin{eqnarray}
\frac{d}{dt} c_v(t) = - \sum_{v' \ne v} p_{v v'}(t)  \, \exp(i(E_v^{\Omega} - E_{v'}^{\Omega}) t/\hbar) \,  c_{v'}(t) 
\end{eqnarray}
with the couplings
\begin{eqnarray}
p_{v v'}(t)  = \langle f_v^{\Omega}(r)  |  \frac{d}{dt} f_{v'}^{\Omega}(r) \rangle = \int_0^\infty  f_v^{\Omega}(r) \, \frac{d}{dt} f_{v'}^{\Omega}(r) \, dr
\end{eqnarray}
obtained from the wavefunctions $f_v^{\Omega}(r)$.
If we consider a constant ramp $R$ of $\Omega$, we can replace the couplings $p_{v v'}(t)$ by $p_{v v'}(\Omega) \times R$,
with 
\begin{eqnarray}
p_{v v'}(\Omega) =  \langle f_v^{\Omega}(r)  |  \frac{d}{d\Omega} f_{v'}^{\Omega}(r) \rangle
\end{eqnarray}
and
\begin{eqnarray}
R = \frac{d\Omega}{dt}.
\end{eqnarray}
For numerical purposes, we use the fact that 
\begin{eqnarray}
p_{v v'}(\Omega)  =  - \langle f_v^{\Omega}(r)  |  \frac{d U_\text{tot}^{\Omega}(r) / d\Omega}{E_v^{\Omega} - E_{v'}^{\Omega}} | f_{v'}^{\Omega}(r) \rangle
\end{eqnarray}
using the Hellmann-Feynman theorem.
The set of equations is solved numerically using a Runge-Kutta method of order 4,
using 120 points from $\Omega = 2\pi \times 10$~MHz to $\Omega= 2\pi \times 40$~MHz, with steps 
of $2\pi \times 0.25$ MHz to converge the results.
The propability to start in $v=0$ and end up in $v'$ for the $l=0$ curve is given by
\begin{eqnarray}\label{Proba}
P_{0 v'}(t) = |c_{v'}(t)|^2 \,  \exp(- t / \tau^{\Omega}) .
\end{eqnarray} 
In Eq.~\eqref{Proba}, $\tau^{\Omega}$ is a quasi-empirical lifetime due to quenching collisions (including sticky collisions \cite{Croft_PRA_102_033306_2020, Croft_PRA_107_023304_2023,Bause_JPCA_127_729_2023}
or chemical reactions, and excitations to higher rotational states from stimulated absorption of microwave photons).  
The exponential decay is then meant to represent the fact that, during the microwave ramp, molecules that encounter one another may be destroyed rather than follow adiabatically into the FLS.  
One can also compute 
the probability $P_{v 0}(t)$ to start in a state 
$v$ and end up in $v'=0$ for the $l=0$ curve
with the $c_{v}(t)$ coefficients subject to different initial conditions in Eq.~\eqref{TDWF}.
This is given by
\begin{eqnarray}
P_{v 0}(t) = |c_{v'=0}(t)|^2 \,  \exp(- t / \tau^{\Omega}) \, /  \, g_v.
\end{eqnarray} 
By dividing by $g_v = (v+1)(v+2)/2$, 
we take into account the unique $l=0$ contribution 
among the $g_v$ 
degenerate states for a given initial $v$, 
as only the $l=0$ curve offers the possibility to have a FLS in $v'=0$.
Note that in Eq.~\eqref{Proba}, $g_{0} = 1$ and does not appear explicitely.

The collisional lifetime of a state is estimated by
\begin{eqnarray}
\tau^{\Omega} = \frac{1}{\beta^{\Omega} \, n(0)}
\end{eqnarray} 
where $\beta^{\Omega}$ is the quenching 
rate coefficient for the two-body scattering problem in free space 
at a given ${\Omega}$ and $n(0)$ the density of the initial gas.
We take $n(0) = 10^{11}$~molecules/cm$^3$ for the case $\nu = 100$~Hz (typical for a thermal gas)
and $n(0) = 1.3 \ 10^{13}$~molecules/cm$^3$ for the case $\nu = 1$~kHz (typical for a BEC).
For states of positive energies, $\beta^{\Omega}$ is the loss rate coefficient
$\beta^{\Omega}_{(\text{NaRb}) - (\text{NaRb})}$
between two colliding NaRb molecules. 
To get a lower limit of the lifetime, we compute $\beta^{\Omega}$ at an energy of $E_c = k_B \times 1$~nK
for each ${\Omega}$ of the ramp
and assign it to each state $v'$ in Eq.~\eqref{Proba}. 
This corresponds to the worst case scenario to show 
what the global effect of losses could be in an experiment.
$\beta^{\Omega}_{(\text{NaRb}) - (\text{NaRb})}$ can peak up to $10^{-10}$~cm$^3$/s
at the resonance seen Fig.~\ref{FIG-SPAG-SCTL}-b but otherwise remains below $10^{-11}$~cm$^3$/s
due to the shielding.

For the state of negative energy, $\beta^{\Omega}$ is the loss rate coefficient
$\beta^{\Omega}_{(\text{NaRb})_2 - (\text{NaRb})_2}$
between two colliding $(\text{NaRb})_2$ FLS. 
We take a unique universal limit value
$\beta_{(\text{NaRb})_2 - (\text{NaRb})_2} = 4.6 \ 10^{-10}$~cm$^3$/s \cite{Idziaszek_PRL_104_113202_2010} 
using a coefficient $C_6^{(\text{NaRb})_2 - (\text{NaRb})_2} = 4 \times C_6^\text{NaRb-NaRb} = 6099600$~a.u~\cite{Lepers_PRA_88_032709_2013}.
Note that for that case, we divide the density by two as the density of the FLS would be twice as small as
that for the diatomic molecules.
We do not consider the dissociation lifetimes of the FLSs as we estimate that they are longer than
the collisional ones (see next section).

\begin{figure}[h]
\begin{center}
\includegraphics*[width=8.5cm,trim=0cm 0cm 0cm 0cm]{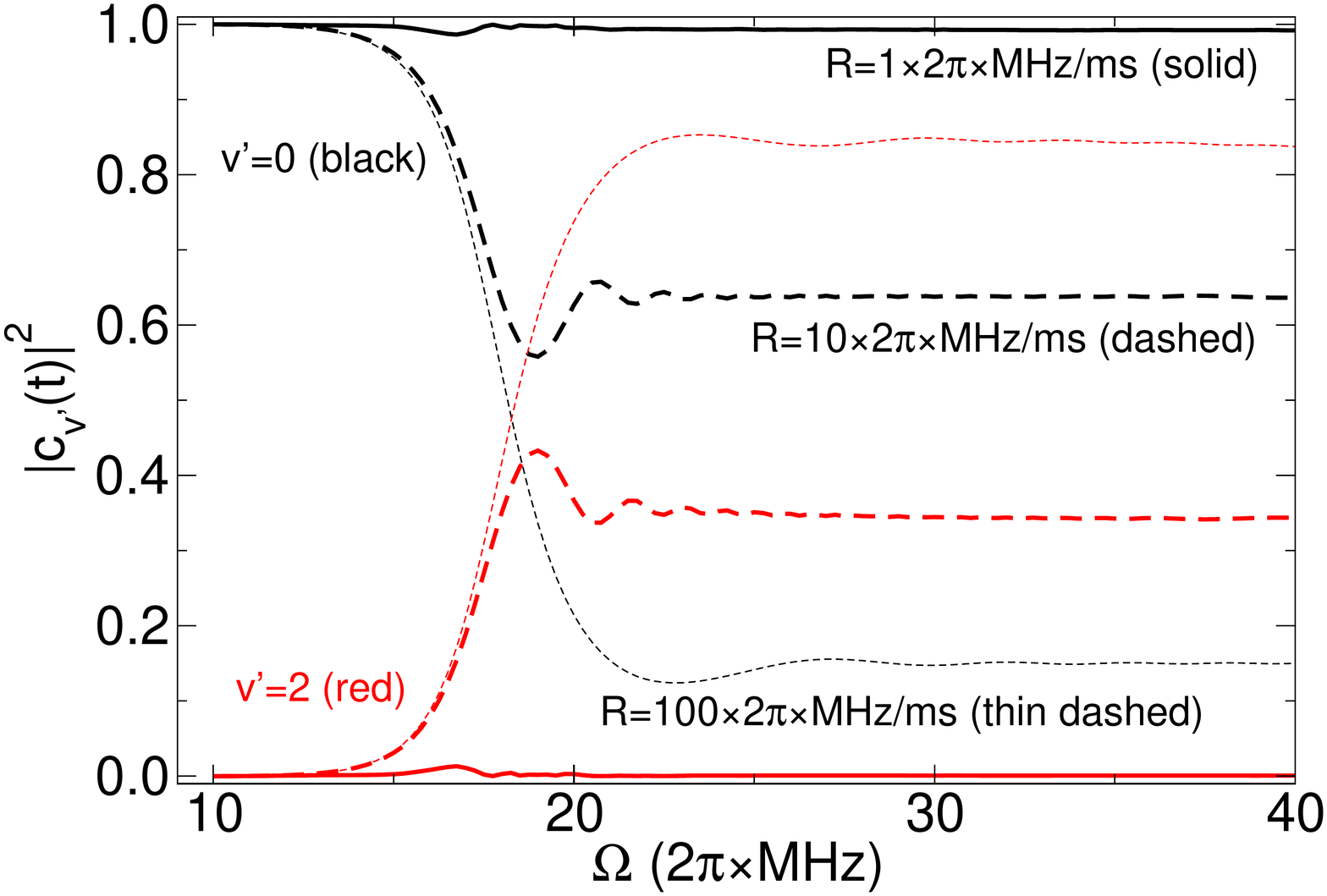} 
\caption{
Probability $|c_{v'}(t)|^2$ 
to start in the state $v=0$ at $\Omega= 2\pi \times 10$~MHz and finish
in the state $v'=0$ (black curves) and $v'=2$ (red curves) at $\Omega= 2\pi \times 40$~MHz
for different ramps $R$, for $\nu = 100$~Hz.
By dividing $\Omega$ by $R$ 
on the abscissa, we get the corresponding time $t$ of the ramp (with the initial time at $\Omega= 2\pi \times 10$~MHz). 
}
\label{FIG-RAMP-TIME}
\end{center}
\end{figure}

For a trap frequency of $\nu=100$~Hz, we found 
$\tau^{\Omega} \simeq 43 $~ms which is longer than the duration of 8.8 ms
at the optimal ramp of $R \simeq 2\pi \times$ 3.4 MHz/ms.
For a trap frequency of $\nu'=1$~kHz, we found 
$\tau^{\Omega} \simeq 0.3 $~ms which is also longer than the duration of 0.11 ms
at $R \simeq 2\pi \times$ 260 MHz/ms.
We show 
in Fig.~\ref{FIG-RAMP-TIME}
the probability $|c_{v'}(t)|^2$ 
for $v=0$ and $v'=0,2$ as a function of time for different ramps $R$.
For each ramp, we take the value at the end of the process and use Eq.~\eqref{Proba}
to report it in Fig.~\ref{FIG-RAMP}.

\subsection{Dissociation of the FLS from an adiabatic distorted-wave Born approximation}

\begin{figure}[h]
\begin{center}
\includegraphics*[width=8.5cm,trim=0cm 0cm 0cm 0cm]{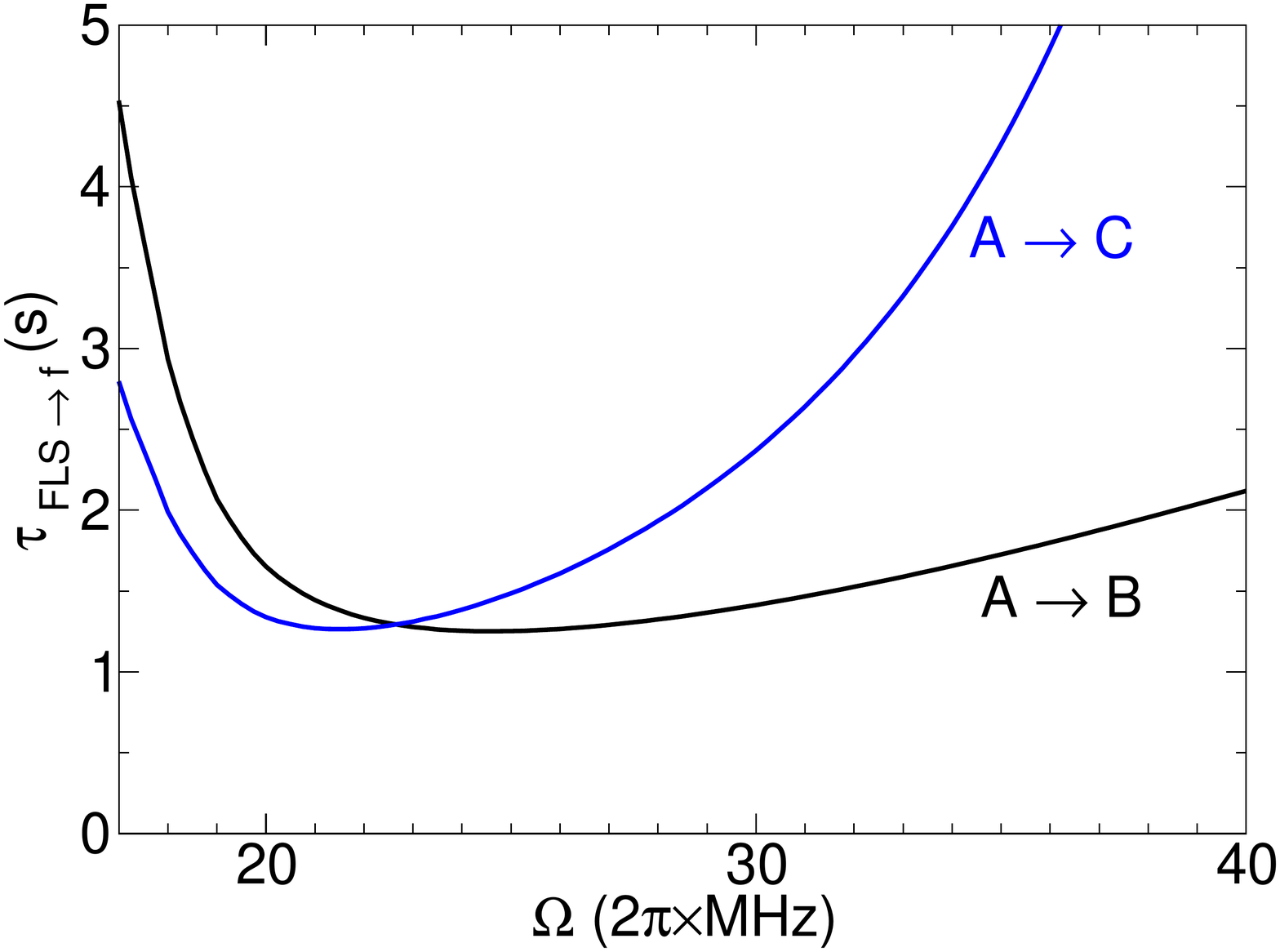} 
\caption{
Decay lifetime $\tau$ as a function of $\Omega$
for the FLS in channel A 
to dissociate in the scattering state of channel B 
or channel C (see Fig.~\ref{FIG-SPAG-SCTL}).
We retain the lowest value for each calculation.
}
\label{FIG-TAU}
\end{center}
\end{figure}

In this section we consider 
the possibility for the FLS carried in the red curve shown
in Fig.~\ref{FIG-SPAG-SCTL}-a, corresponding to an entrance channel with initial $l=0$,
to decay into a lower scattering channel with a final value $l$, 
shown as black curves on the same figure.
The general Hamiltonian is $H(r) = -(\hbar^2/2 \mu) d^2/dr^2 + V_\text{eff}$ with the potential 
$V_\text{eff} = \hbar^2 \, l(l+1) / 2 \mu r^2 + V$ including the centrifugal term. 
$V$ is the potential energy
and it depends on $\vec{r}$ but also on the internal coordinates $\vec{\rho}_1, \vec{\rho}_2$
of the molecules.
$V_\text{eff}$ is diagonalized at each $r$
\begin{eqnarray}
V_\text{eff} \, \phi_i  = U_i(r) \, \phi_i
\end{eqnarray}
to give the adiabatic energies $U_i(r)$ (shown in Fig.~\ref{FIG-SPAG-SCTL}-a)
and functions $\phi_i(\hat{r}, \vec{\rho}_1, \vec{\rho}_2;r)$,
which now depend parametrically on $r$.
The time-independent wavefunction of the full coupled system
is expanded in the adiabatic functions
\begin{eqnarray}
\psi(\vec{r}, \vec{\rho}_1, \vec{\rho}_2) = \sum_i F_i(r) \, \phi_i(\hat{r}, \vec{\rho}_1, \vec{\rho}_2;r).
\end{eqnarray}
The coupled Schr\"odinger equations become
\begin{multline}
- \frac{ \hbar^2 }{ 2 \mu  } \sum_i \left[ \frac{ d^2 F_i }{ dr^2 } \phi_i 
+ 2 \frac{ d F_i }{ dr }   \frac{ d \phi_i }{ dr } + F_i \frac{ d^2 \phi_i }{ dr^2 } \right] \\
+  \sum_i U_i F_i \phi_i = E \sum_i F_i \phi_i.
\end{multline}
We project these equations onto a single adiabatic channel $\phi_j$, where the projection is over all coordinates except $r$.  
Here we retain only two channels, namely $\phi_\text{A}$ for the channel carrying the FLS 
and $\phi_\text{B}$ for the channel into which the FLS will decay.
These two channels are indicated in Fig.~\ref{FIG-SPAG-SCTL}-a.
We moreover ignore the second derivatives $d^2 \phi_i / dr^2$.  
The coupled channel equations become
\begin{eqnarray}
- \frac{ \hbar^2 }{ 2 \mu } \frac{ d^2F_\text{A} }{ dr^2 } + U_\text{A} F_\text{A} - E F_\text{A} = \frac{ \hbar^2 }{ \mu } P_{\text{AB}} \frac{ dF_\text{B} }{ dr }
\nonumber \\
- \frac{ \hbar^2 }{ 2 \mu } \frac{ d^2F_\text{B} }{ dr^2 } + U_\text{B} F_\text{B} - E F_\text{B} = \frac{ \hbar^2 }{ \mu } P_{\text{BA}} \frac{ dF_\text{A} }{ dr }.
\end{eqnarray}
Here $P_{\text{AB}} = \langle \phi_\text{A} | \frac{d}{dr} | \phi_\text{B} \rangle$ are the non-adiabatic couplings and have vanishing diagonal elements. The integration is over all but the radial coordinate.
We assume the field-linked state is a bound state of channel A, near energy $E$; and that it will dissociate into channel B, with kinetic energy $E-E_\text{B}$, where $E_\text{B}$ is the threshold energy of channel B.  
Ignoring the non-adiabatic coupling on the right of the second equation, 
we define reference functions via
\begin{eqnarray}
\left( - \frac{ \hbar^2 }{ 2 \mu } \frac{ d^2 }{ dr^2 } + U_\text{B}  - E \right) 
\left\{  \begin{array}{c} f_\text{B} \\ g_\text{B} \end{array} \right\} = 0
\end{eqnarray}
with boundary conditions for large $r$
\begin{eqnarray}\label{fB}
f_\text{B} &\rightarrow&  \sqrt{ \frac{ 2 \mu }{ \pi \hbar^2 k } } \sin(kr - l\pi/2 + \delta_l) \nonumber \\
g_\text{B} &\rightarrow&  - \sqrt{ \frac{ 2 \mu }{ \pi \hbar^2 k } } \cos(kr - l\pi/2 + \delta_l) 
\end{eqnarray}
where $k = \sqrt{ 2 \mu (E - E_\text{B})/ \hbar^2 }$.  
These functions are energy normalized
\begin{eqnarray}
\langle f_\text{B}(E) | f_\text{B}(E^{\prime}) \rangle = \delta (E - E^{\prime})
\end{eqnarray}
and define a Green function~\cite{Friedrich_Book_2005}
\begin{eqnarray}
G(r,r^{\prime}) = - \pi f_\text{B}(r_<) g_\text{B}(r_>).
\end{eqnarray}
The wavefunction in channel B (for a given partial wave $l$)
is then given in the Born approximation by
\begin{eqnarray}
F_\text{B}(r) = f_\text{B}(r) + \int_0^{\infty} dr^{\prime} G(r,r^{\prime}) \frac{ \hbar^2 }{ \mu } P_{\text{BA}}
\frac{ dF_\text{A} }{ dr } (r^{\prime})
\end{eqnarray}
and in the limit $r \rightarrow \infty$ this becomes
\begin{multline}
F_\text{B} \rightarrow \sqrt{ \frac{ 2 \mu }{ \pi \hbar^2 k } } 
\big\{ \sin( kr - l\pi/2 + \delta_l) \\
+ K \cos( kr - l\pi/2 + \delta_l) \big\} .
\end{multline}
We can identify the $K$-matrix
\begin{eqnarray}
K =   \pi \frac{ \hbar^2 }{ \mu } \int_0^{\infty} dr f_\text{B}  P_{\text{BA}}
\frac{ dF_\text{A} }{ dr }.   
\end{eqnarray}
As for $F_\text{A}$, it is well approximated by a bound state at energy $E_0$
\begin{eqnarray}
F_\text{A} = N f_0
\end{eqnarray}
where $f_0$ is the wavefunction of the FLS (using the notation of the paper), normalized conventionally
\begin{eqnarray}
\int_0^{\infty} dr f_0^2(r) = 1
\end{eqnarray}
and where $N$ is a leftover normalization in the context of the coupled channels.  
Then the equation for $F_\text{A}$ reads
\begin{eqnarray}
(E_0-E) N f_0 = \frac{ \hbar^2 }{ \mu } P_{\text{AB}} \frac{ dF_\text{B} }{ dr } .
\end{eqnarray}
Multiplying this by $f_0$ and integrating over $r$ gives
\begin{eqnarray}
(E_0 - E) N =  \frac{ \hbar^2 }{ \mu } \int_0^{\infty} dr f_0 P_{\text{AB}} \frac{ dF_\text{B} }{ dr } \nonumber \\
\approx  \frac{ \hbar^2 }{ \mu } \int_0^{\infty} dr f_0 P_{\text{AB}} \frac{ df_\text{B} }{ dr }.
\end{eqnarray}
Thus $N$ is determined and we have an expression for the $K$-matrix,
\begin{multline}
K =  \pi \left(  \frac{ \hbar^2 }{ \mu } \int_0^{\infty} dr f_\text{B} P_{\text{BA}} \frac{ df_0 }{ dr } \right) N  \\
=  \pi \left(  \frac{ \hbar^2 }{ \mu } \int_0^{\infty} dr f_\text{B} P_{\text{BA}} \frac{ df_0 }{ dr } \right) \\
 \left(  \frac{ \hbar^2 }{ \mu } \frac{ 1 }{E_0 - E}  \int_0^{\infty} dr f_0 P_{\text{AB}} \frac{ df_\text{B} }{ dr } \right).
\end{multline}
As this is a resonant $K$-matrix defining a resonant phase-shift, 
we have \cite{Brandsen_Joachain_Book_2003}
\begin{eqnarray}
K = \tan \delta_{res} = \frac{  \Gamma / 2 }{ E_0 - E }
\end{eqnarray}
and we get the resonance width
\begin{multline}\label{gamma}
\Gamma  = 2 \pi \left(  \frac{ \hbar^2 }{ \mu } \int_0^{\infty} dr f_\text{B} P_{\text{BA}} \frac{ df_0 }{ dr } \right) 
 \left(  \frac{ \hbar^2 }{ \mu }  \int_0^{\infty} dr f_0 P_{\text{AB}} \frac{ df_\text{B} }{ dr } \right) \\
 = ( 4 \hbar^2 / \mu)  \int_0^{\infty} dr 
 \sin(kr - l\pi/2 + \delta_l) P_{\text{BA}} \frac{ df_0 }{ dr }  \\
\times  \int_0^{\infty} dr f_0 P_{\text{AB}} \cos(kr - l\pi/2 + \delta_l)  
\end{multline}
where in the last line we used to a good approximation
the asymptotic form Eq.~\eqref{fB} of $f_\text{B}$.
The absolute value of this width therefore approximates the decay rate of the field-linked state
for which the decay lifetime is given by
\begin{eqnarray}\label{Tau-ADWBA}
\tau = \frac{\hbar}{|{ \Gamma }|} .
\end{eqnarray}
The lifetime will depend of the phase-shift $\delta_l$
appearing in Eq.~\eqref{gamma}
when the FLS dissociate in the scattering channel.
One can compute this value using our close-coupling calculation,
however this value will not be converged as it will depend
on the minimum radial distance of the
wavefunction taken for solving the equations.
Instead, we scan different values of $\delta_l$
from 0 to $\pi$
and we retain the lowest corresponding lifetime.
This is shown in Fig.~\ref{FIG-TAU} for the dissociation of the FLS 
to the channel B or channel C, 
indicated in Fig.~\ref{FIG-SPAG-SCTL}. Dissociation to the other channels takes longer.
We evaluated $\Gamma$ and $\tau$ at $E = E_0$.
We can see that the shortest decay lifetime is around 1.25 s, which is 
longer than the collisional lifetime of the FLS that is on the order of ms.
As such we can neglect dissociation of the FLS during the ramp process.

To compare with the original static field-linked state (sFLS)
where dissociation lifetimes of 3 ns 
were predicted~\cite{Avdeenkov_PRA_69_012710_2004},
microwave field-linked states (mFLS) seem to be longer-lived.
This can be qualitatively explained by the fact that the mFLS occurs at much larger
radial distance than the original one, where dipolar couplings 
are smaller.
From a rough approximation,  
if we take an infinite square-potential function for the FLS and its derivative, $f_0 \propto (\Delta r)^{-1/2}$ 
and $df_0/dr \propto {\Delta r}^{-3/2}$ where $\Delta r$ is the range of the FLS well.
From the Hellmann--Feynman theorem, we also have $P_{\text{AB}}= \langle \phi_A | V(r') |  
\phi_B \rangle / [(r'-r) (U_\text{B}(r) - U_\text{A}(r))]$ where $\phi_\text{A,B}$ are evaluated at $r$
and $V$ evaluated at a nearby point $r' \simeq r$.
As $V(r') \simeq V(r) \propto d^2 / r^{3}$ from the dipole-dipole interaction with dipole $d$
and as $(r'-r)$ will scale the same way as $dr$ in the integrals, we see that 
\begin{align}
\Gamma &\propto d^4 / (r^{6} \, \Delta r^{2}) & \tau &\propto r^{6} \, \Delta r^{2} / d^4.
\end{align}
If we take the values $d_\text{sFLS}$ = 1.67 D, 
$r_\text{sFLS} \simeq 80 \, a_0$ and $\Delta r_\text{sFLS} \simeq 50 \ a_0$
from \cite{Avdeenkov_PRA_69_012710_2004} to compare with $d_\text{mFLS}$ = 3.2 D,
$r_\text{mFLS} \simeq 1300 \ a_0$ and $\Delta r_\text{mFLS} \simeq 2000 \ a_0$ from this work, we see that $\tau_\text{mFLS} / \tau_\text{sFLS} \simeq 2 \, 10^{9}$. 
Taking $\tau_\text{sFLS} = 3$~ns \cite{Avdeenkov_PRA_69_012710_2004}, this makes a rescaled value of $\tau_\text{mFLS} = 6$~s, which is on the same order with what we computed using Eq.~\eqref{gamma} with a value of 1.25~s.

\bibliography{../../../BIBLIOGRAPHY/bibliography}

\end{document}